\documentclass[english,prb,twocolumn]{revtex4}
\pdfoutput=1
\usepackage[T1]{fontenc}
\usepackage[latin9]{inputenc}
\usepackage{amsthm}
\usepackage{amsmath}
\usepackage{amssymb}
\usepackage{graphicx}

\makeatletter
\@ifundefined{textcolor}{}
{%
 \definecolor{BLACK}{gray}{0}
 \definecolor{WHITE}{gray}{1}
 \definecolor{RED}{rgb}{1,0,0}
 \definecolor{GREEN}{rgb}{0,1,0}
 \definecolor{BLUE}{rgb}{0,0,1}
 \definecolor{CYAN}{cmyk}{1,0,0,0}
 \definecolor{MAGENTA}{cmyk}{0,1,0,0}
 \definecolor{YELLOW}{cmyk}{0,0,1,0}
}

\makeatother

\usepackage{babel}
\begin{document}

\title{Strong driving of a single spin using arbitrarily polarized fields}

\author{P. London,$^{1,*}$ P. Balasubramanian,$^{2}$ B. Naydenov,$^{2}$
L. P. McGuinness,$^{2,\dagger}$ and F. Jelezko$^{2}$}

\affiliation{$^{1}$Department of Physics, Technion, Israel Institute of Technology,
Haifa 32000, Israel}

\affiliation{$^{2}$Institut f�r Quantenoptik, Universitat Ulm, 89073 Ulm, Germany}
\begin{abstract}
The strong driving regime occurs when a quantum two-level system is
driven with an external field whose amplitude is greater or equal
to the energy splitting between the system's states, and is typically
identified with the breaking of the rotating wave approximation (RWA).
We report an experimental study, in which the spin of a single nitrogen-vacancy
(NV) center in diamond is strongly driven with microwave (MW) fields
of arbitrary polarization. We measure the NV center spin dynamics
beyond the RWA, and characterize the limitations of this technique
for generating high-fidelity quantum gates. Using circularly polarized
MW fields, the NV spin can be harmonically driven in its rotating
frame regardless of the field amplitude, thus allowing rotations around
arbitrary axes. Our approach can effectively remove the RWA limit
in quantum-sensing schemes, and assist in increasing the number of
operations in QIP protocols.
\end{abstract}
\maketitle

\section{Introduction}

The nitrogen-vacancy (NV) center is one of the leading platforms for
QIP applications \cite{LaddNature2010,WaldherrNature2014}, and room-temperature
quantum metrology \cite{MaminScience2013,StaudacherScience2013}.
Additionally, it serves as a probe for the classical and quantum dynamics
of a mesoscopic bath of spins \cite{deLangeScRep2012,ReinhardPRL2012}.
These applications provide great motivation for controlling, and specifically
shortening the manipulation duration of the spin. In QIP, shortening
of the gate duration allows an increase in the number of quantum-gates
applied during the coherence time, $T_{2}$, and thus scales up the
computational performance\cite{TaminiauNatNano2014}. Ultimately,
the elementary gate duration defines the processing clock-speed \cite{PressNature2008};
for systems on parity for the number-of-operations figure of merit,
it distinguishes between ``slow'' systems, such as cold trapped
atoms or nuclear spins, and ``fast'' systems such as semiconductor
quantum dots and superconducting flux qubits. In quantum metrology,
designed rotations of qubits are used to map the signal (the phase
between eigenstates) to a measurable population difference. Specifically,
in dynamical decoupling (DD) based quantum sensing schemes, the qubit
may be driven continuously or pulsed at intervals, allowing suppression
of noise sources with a slower spectrum than the driving speed/interpulse
spacing \cite{Taylor2008,TaminiauPRL2012,KolkowitzPRL2012}. Thus,
the maximum driving speed or pulse duration places an upper bound
on the ability to shift the sensing frequencies higher and away from
the dominant low frequency noise \cite{romach_arxiv}, and limits
the bandwidth of these schemes.

For these goals, and crucially in room temperature applications, the
NV spin is usually manipulated with an oscillatory microwave (MW)
field \textbf{$B_{x}\left(t\right)=B_{1}\cos\left(\omega t\right)$},
(where $B_{1}$, $\omega$ are the field amplitude and frequency respectively),
resonant with the energy splitting of the spin $\hbar\omega_{L}$,
i.e. $\omega=\omega_{L}.$ Then, in a frame rotating with the MW field,
the spin is driven by a constant magnetic field $B_{+}=B{}_{1}/2$
(co-rotating field), and an additional rotating field $B_{-}=\left(B{}_{1}/2\right)e^{2i\omega t}$
(counter-rotating field). As long as $\gamma B_{1}$ is small compared
to $\omega$ (where $\gamma$ is the magnetic moment of the spin),
the counter-rotating field can be neglected, an approximation known
as the rotating wave approximation (RWA). In this regime, the gate
time depends linearly on the inverse of applied magnetic field amplitude
$B_{1}$. 

However, when the driving amplitude is increased so that $\gamma B_{1}$
becomes comparable to the spin's Larmor frequency $\omega_{L}$, the
dynamics manifest complexities due to an interplay of the two fields:
the gate fidelity degrades, the rotation (gate) time no longer scales
linearly with $1/B_{1},$ and the dynamics show pronounced sensitivity
to the phase of $B_{x}\left(t\right)$ with respect to the pulse edges
\cite{FuchsScience2009}. In this regime, known as the strong driving
regime, various solutions to regaining control of the system dynamics
have been proposed, including anharmonic pulses \cite{AvinadavArXiv2014,ScheuerArXiv2014},
Landau-Zener assisted transitions \cite{ZhouPRL2014}, and transitions
through an ancillary level in a $\Lambda$-type configuration \cite{KodiranoPRB2012}. 

Here we tackle the strong driving problem using an approach discussed
in an early work of Bloch and Siegert \cite{BlochPhysRev1940}. A
spin subjected to two orthogonal, resonant MW fields $B_{x}=B_{1}\cos(\omega_{L}t)$
and $B_{y}=B_{1}\cos(\omega_{L}t+\phi)$ will rotate harmonically
if $\phi=\pm\pi/2$. Under this condition, the two orthogonal fields
can equivalently be described as circularly polarized MW radiation.
When the radiation polarization coincides with the spin transition
(i.e. when the angular momentum of the radiation field matches the
change in spin number), manipulation with a field solely co-rotating
with the spin occurs, leading to full contrast rotations. The other
case, namely driving a transition with with the counter-rotating field
(of opposite handedness/polarization), can be viewed as a driving
field with a $2\omega_{L}$ detuning. Only in the strong driving regime,
may rotation of the spin occur, albeit with degraded contrast. Recently,
this aproach was demonstrated with an ensemble of $^{1}$H nuclear
spins ($I=1/2$) in an ultra-low field NMR setup \cite{ShimJMR2014}.
Here, we investigate this approach using experiments on a single NV
center, an electronic spin in diamond with total spin $S=1$. For
an $S=1$ system, it is possible to address more than one transition
spectrally, enabling polarization selective transitions \cite{AlegrePRB2007}.
We drive the NV center with arbitrarily polarized MW radiation, address
one of the two-level systems, and study its dynamics in the strong
driving regime, namely, when the Rabi frequency is larger than the
Larmor frequency.

The paper is organized as follows. In Section II we describe the experimental
setup, and present a theoretical description of the general Hamiltonian
of the NV $S=1$ ground-state under two MW fields. Section III discusses
the dynamics in low magnetic field, characterized by selective excitation
within a dense spectrum of resonances. In Section IV we experimentally
demonstrate the strong driving regime for various polarizations, and
compare between the dynamics under linear and circular polarizatons.
In Section V we discuss the results and elaborate on the effect of
an axial MW component on the NV dynamics, i.e a MW field applied parallel
to the NV dipole axis.

\section{Experimental setup for polarized MW radiation}

\begin{figure}
\includegraphics[bb=0bp 0bp 380bp 540bp,clip,width=1\columnwidth]{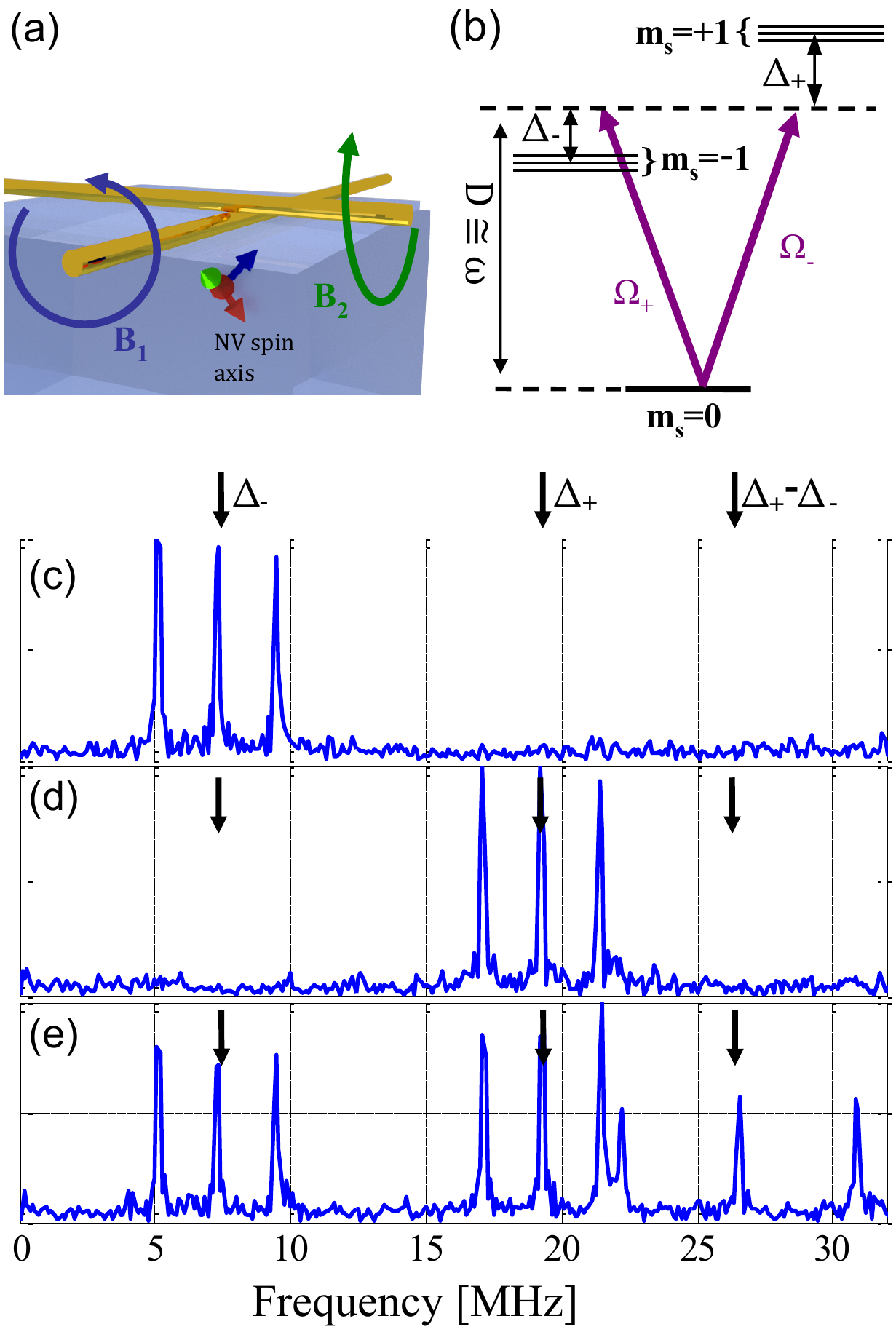}

\caption{Experimental setup and broadband magnetometry with polarized fields
(a) Microwaves transmitted by two crossed wires placed on the diamond
surface create orthogonal magnetic fields, and for certain NV positions
allow an arbitrary MW field. (b) Energy scheme of the NV ground-state
at low magnetic field. The notations are assgined in the text. (c-e)
Experimental spectra of the Ramsey signals produced with MW $\pi/2$
pulses of right-handed circular polarization, left-handed circular
polarization, and linear polarization respectively. The black arrows
in (c)-(e) mark the frequencies $\Delta_{\pm}$, and $\Delta_{+}-\Delta_{-}$
at the applied magnetic field. \label{fig:SetupAndRamsey}}
\end{figure}

The experiments were conducted at room temperature, with single NV
centers in a type IIa diamond with (100) surface. To excite NV centers
with arbitrarily polarized MW pulses \cite{AlegrePRB2007} of short
duration, we designed a low-Q MW antenna. The antenna comprises of
two thin copper wires in a cross-configuration, stretched over the
diamond surface (Fig. \ref{fig:SetupAndRamsey}a). The wires were
connected to two independent MW sources, switches, and amplifiers,
and were phase-locked to each other. Alternatively, one can apply
the fields through an arbitrary waveform generator to gain full control
over the MW parameters. With this setup we were able to manipulate
individual NV centers, located at distances of $\sim$10-50$\mu$m
from the wire crossing, with Rabi frequencies up to 100 MHz. We have
found that the position of the NV center with respect to the wires
affects the driving performance (See Sec.V). The ideal scenario is
illustrated in Fig. \ref{fig:SetupAndRamsey}a, where the two fields
and the NV axis form an orthogonal system.

The Hamiltonian of the NV center spin, $\mathbf{S}$, in the presence
of two orthogonal driving fields $B_{x},B_{y}$ of equal magnitude,
and a constant external magnetic field $B_{ext}$, can be written
as
\begin{equation}
H=DS_{z}^{2}-\gamma B_{ext}S_{z}+\Omega e^{i\left(\omega t+\phi_{g}\right)}\hat{\varepsilon}\cdot\mathbf{S}+\mathrm{h.c.},\label{eq:BasicHamiltonian}
\end{equation}
where $D=\left(2\pi\right)$2.87 GHz is the zero-field splitting,
$\gamma=\left(2\pi\right)$2.8 MHz/G is the NV magnetic moment, $\hat{\varepsilon}=\left(1,e^{i\phi},0\right)$
defines the MW polarization and h.c. stands for hermitian conjugate.
Here, $\omega$, $\Omega$, and $\phi$ are the MW frequency, the
NV Rabi frequency $\left(\Omega=\gamma B_{x}=\gamma B_{y}\right)$,
and relative MW phase, respectively. The phase $\phi_{g}$ is a \textbf{global
phase} shared by both fields. In the rotating frame, Eq. (\ref{eq:BasicHamiltonian})
is rewritten as (See Apendix A)
\begin{eqnarray}
H' & = & \left[\begin{array}{ccc}
\Delta_{-} & 0 & 0\\
0 & 0 & 0\\
0 & 0 & \Delta_{+}
\end{array}\right]+\frac{\Omega}{\sqrt{2}}\left[\begin{array}{ccc}
0 & \varepsilon_{-} & 0\\
\varepsilon_{-}^{*} & 0 & \varepsilon_{+}\\
0 & \varepsilon_{+}^{*} & 0
\end{array}\right]\nonumber \\
 & + & \frac{\Omega}{\sqrt{2}}\left[\begin{array}{ccc}
0 & \varepsilon_{+}e^{-2i\omega t} & 0\\
\varepsilon_{+}^{*}e^{2i\omega t} & 0 & \varepsilon_{-}e^{2i\omega t}\\
0 & \varepsilon_{-}^{*}e^{-2i\omega t} & 0
\end{array}\right]\label{eq:FullHamiltonian}
\end{eqnarray}
where $\varepsilon_{\pm}=e^{\mp i\phi_{g}}\left(1-ie^{\mp i\phi}\right)/2$.
Here $\omega_{L}^{\pm}=D\pm\gamma B_{ext}$ are the transition frequencies,
and $\Delta_{\pm}=\omega_{L}^{\pm}-\omega$ their detuning from the
microwave frequency (Fig. \ref{fig:SetupAndRamsey}b). The second
and the third terms of Eq.(\ref{eq:FullHamiltonian}) represent the
co-rotating and counter-rotating terms, respectively. Note that Eqs.(\ref{eq:BasicHamiltonian},\ref{eq:FullHamiltonian})
hold for arbitrarily (elliptically) polarized fields, however we assume
that the z-component of the MW field is zero. We refine this treatment
in Section V, when discussing the influence of a MW field with non-zero
axial component. 

Next we discuss the dynamics of Eq.(\ref{eq:FullHamiltonian}) in
the `low field' case, where the transitions are nearly degenerate
with respect to the Rabi frequency, but the RWA is applicable ($\Delta_{\pm}\ll\Omega\ll\omega$).
Then, we discuss the dynamics in the `high-field' case where the transitions
are well separated and the Rabi frequency exceeds the transition frequency
($\Delta_{-}\ll\omega\leq\Omega$), allowing investigation of a two-level
system driven beyond the RWA limit.

\section{Selective excitation with polarized fields}

To characterize the performance of the MW structure we perfomed broadband
Ramsey magnetometry at low magnetic field, where the $\left|-1\right\rangle $
and $\left|+1\right\rangle $ states are nearly degenerate \cite{CaoChinPhysLett1997}.
For low amplitude driving $\left(\Omega\ll\omega\right)$ one may
use the RWA, i.e. assume that the $\varepsilon_{\pm}e^{\pm2i\omega t}$
components oscillate many times during the rotation of the spin, and
thus are averaged to zero. Eq.(\ref{eq:FullHamiltonian}) then becomes
\begin{equation}
H^{RWA}=\left[\begin{array}{ccc}
\Delta_{-} & \left(\Omega/\sqrt{2}\right)\varepsilon_{-} & 0\\
\left(\Omega/\sqrt{2}\right)\varepsilon_{-}^{*} & 0 & \left(\Omega/\sqrt{2}\right)\varepsilon_{+}\\
0 & \left(\Omega/\sqrt{2}\right)\varepsilon_{+}^{*} & \Delta_{+}
\end{array}\right].
\end{equation}
Here we see that $\left(\Omega/\sqrt{2}\right)\varepsilon_{+}$ drives
the $\left|0\right\rangle \longleftrightarrow\left|+1\right\rangle $
transition, and $\left(\Omega/\sqrt{2}\right)\varepsilon_{-}$ drives
the $\left|0\right\rangle \longleftrightarrow\left|-1\right\rangle $
transition.

The NV is first optically pumped to the $\left|0\right\rangle $ state.
Then, a MW $\pi/2$-pulse with arbitrary polarization (arbitrary $\phi$)
manipulates the NV spin to the state $\left|\psi\right\rangle =\left(1/\sqrt{2}\right)\left[\left|0\right\rangle +\varepsilon_{+}\left|1\right\rangle +\varepsilon_{-}^{*}\left|-1\right\rangle \right]$.
This state can be obtained with the evolution operator $U=\exp\left(iH^{RWA}t\right)$,
for rotation time $t$ satisfying $\left(t\cdot\Omega/\sqrt{2}\right)=\pi/2$,
in the limit $\Omega\gg\Delta_{\pm}$. Then, after a free-evolution
time, $\tau$, the state becomes $\left|\psi\right\rangle =\left(1/\sqrt{2}\right)\left[\left|0\right\rangle +\varepsilon_{+}e^{i\Delta_{+}\tau}\left|1\right\rangle +\varepsilon_{-}^{*}e^{-i\Delta_{-}\tau}\left|-1\right\rangle \right]$,
and an additional $\pi/2$-pulse with the same polarization gives
the final probability to be in the $\left|0\right\rangle $ state,
as
\begin{eqnarray}
P_{0}\left(\tau\right) & = & \frac{1}{2}\left[1-\left|\varepsilon_{-}\right|^{2}\sin\left(\Delta_{-}\tau\right)-\left|\varepsilon_{+}\right|^{2}\sin\left(\Delta_{+}\tau\right)\right.\nonumber \\
 & - & \left.\left|\varepsilon_{-}\right|^{2}\left|\varepsilon_{+}\right|^{2}\left(\cos\left[\left(\Delta_{+}-\Delta_{-}\right)\tau\right]-1\right)\right]\label{eq:BroadBandRamsey}
\end{eqnarray}
The second and third terms oscillate at the microwave frequency detuning
from the $\left|0\right\rangle \longleftrightarrow\left|-1\right\rangle $
and $\left|0\right\rangle \longleftrightarrow\left|+1\right\rangle $
transitions, respectively. The last term oscillates at the frequency
separation between the $\left|\pm1\right\rangle $ states, and is
detuning independent. Using Fourier analysis of $P_{0}\left(\tau\right)$,
one can infer the polarization parameters ($\left|\varepsilon_{\pm}\right|^{2}$)
directly, by observing the intensity of each frequency component. 

In the experiments, a static axial magnetic field of 4.6 Gauss was
used to split the $\left|\pm1\right\rangle $ states by 26 MHz, and
`hard' $\pi/2$-pulses which efficiently excited both transitions
were applied ($\Omega=\left(2\pi\right)$114MHz, note that $\Omega$
remained still much smaller than $\omega_{L}\simeq\left(2\pi\right)$3
GHz). By varying the relative phase between the wires, various polarizations
could be engineered; left-handed circular polarization (driving the
$\left|0\right\rangle \rightarrow\left|-1\right\rangle $ transition,
Fig. \ref{fig:SetupAndRamsey}c), right-handed circular polarization
(driving the $\left|0\right\rangle \rightarrow\left|+1\right\rangle $
transition, Fig. \ref{fig:SetupAndRamsey}d), and linear polarization
(Fig. \ref{fig:SetupAndRamsey}e). In all spectra there is an additional
$\left(2\pi\right)2.16$MHz splitting due to hyperfine interaction
with the NV host nitrogen nuclear spin. From the relative amplitudes
in the spectral footprint, we deduce that $\left|\varepsilon_{-}\right|^{2}=0.98,0.03,0.47$(with
an error of $\pm0.05$) for Figs.\ref{fig:SetupAndRamsey}c,d and
e, respectively.

\begin{figure*}
\includegraphics[bb=0bp 50bp 960bp 540bp,clip,width=1\textwidth]{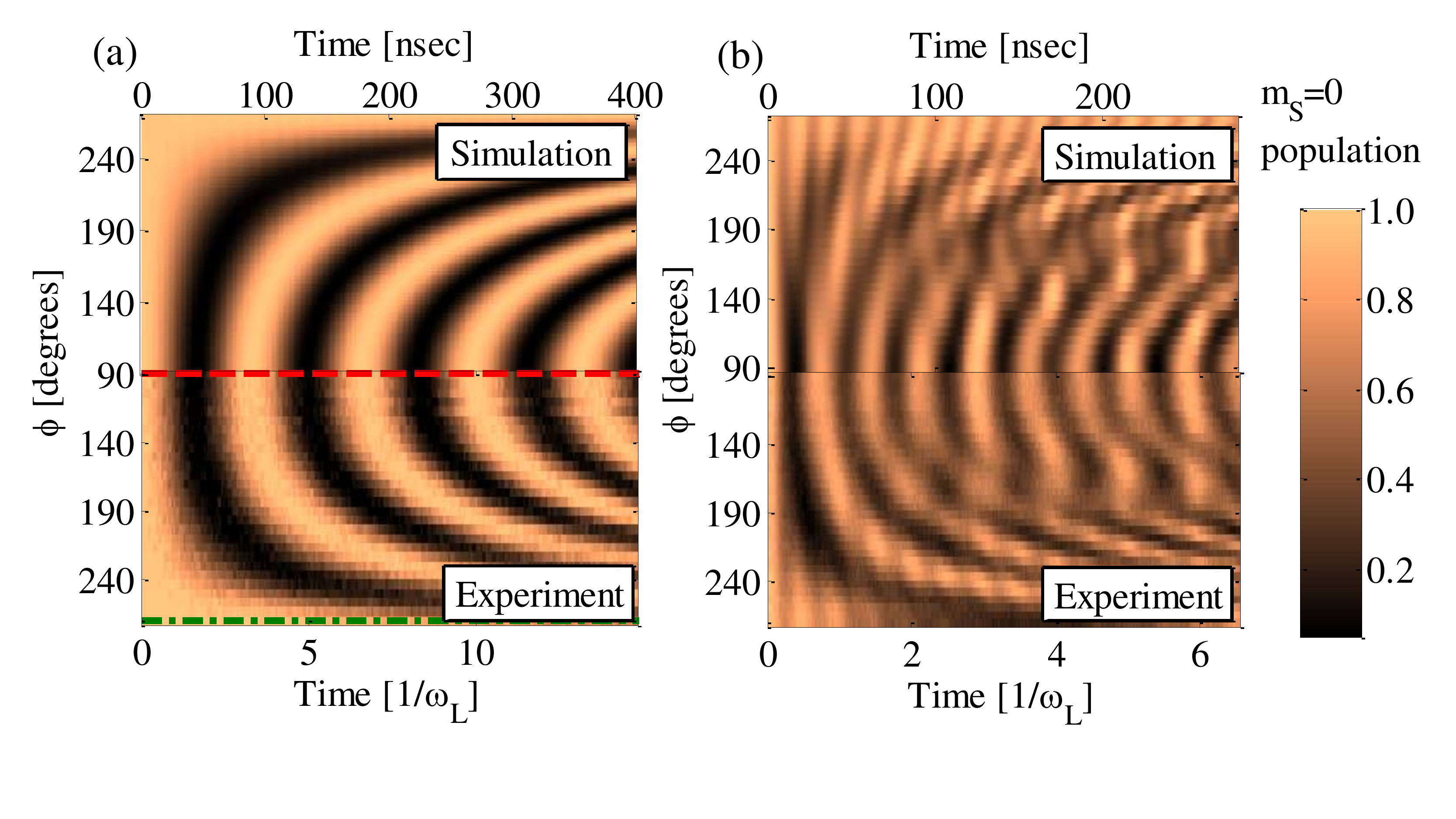}

\caption{Rabi oscillations with MW fields of various polarizations. Simulated
(upper panels) and experimental (lower panels) Rabi oscillations with
various phase difference between the wires, in (a) the weak driving
and (b) the strong driving regimes. In the simulation, an average
of 300 initial phases $\phi_{g}$ were used.\label{fig:RabiCircular}}
\end{figure*}

\section{Strong driving with arbitrary polarizations}

After characterizing the MW polarization, we experimentally investigated
the strong driving regime for different applied polarizations. A two-level
system (TLS) was prepared using a high axial magnetic field of $\sim$0.1
mT. At this field one finds $\omega_{L}^{-}=\left(2\pi\right)30$MHz,
and $\omega_{L}^{+}=\left(2\pi\right)5710$MHz. For a field resonant
with $\omega_{L}^{-}$, the far-detuned higher energy transition can
be neglected, and the reduced Hamiltonian of the two-level system
$\left|0\right\rangle ,\left|-1\right\rangle $ {[}derived from Eq.(\ref{eq:FullHamiltonian}){]}
is
\begin{equation}
H=\frac{\Omega}{\sqrt{2}}\left[\begin{array}{cc}
0 & \varepsilon_{-}+\varepsilon_{+}e^{-2i\omega_{L}^{-}t}\\
\varepsilon_{-}^{*}+\varepsilon_{+}^{*}e^{2i\omega_{L}^{-}t} & 0
\end{array}\right]\label{eq:StrongDrivingHamiltonian}
\end{equation}
where $\left(\Omega/\sqrt{2}\right)\varepsilon_{-}$ is the co-rotating
component of the MW field, and $\left(\Omega/\sqrt{2}\right)\varepsilon_{+}e^{-2i\omega_{L}^{-}t}$
is the counter-rotating component. Specifically, for $\phi=\pi/2$,
one obtains the Hamiltonian $H=\left(\Omega/\sqrt{2}\right)\sigma_{x}$
where $\sigma_{x}$ is the Pauli matrix, and the prefactor $\sqrt{2}$
is a remnant of the $S=1$ nature of the NV system (The spin interacts
stronger than a true TLS). This Hamiltonian is exact, and independent
of the driving field magnitude, even for $\Omega\geq\omega_{L}^{-}$,
i.e. beyond the RWA limit. The dynamics derived from this Hamiltonian
are harmonic oscillations with Rabi frequency $\Omega$. We note that
our experiments were conducted with Rabi frequencies on the order
of tens of MHz, but in principle could be performed at the GHz regime
with the proper hardware \cite{FuchsScience2009}. The approximation
leading from Eq.(\ref{eq:FullHamiltonian}) to Eq.(\ref{eq:StrongDrivingHamiltonian})
breaks only at $\Omega\sim2D\simeq\left(2\pi\right)6$GHz. For these
values parasitic excitations to the $\left|+1\right\rangle $ state
will interfere with the dynamics.

\subsection{Optimization of the relative phase $\phi$}

As described above, experimental control of the MW polarization is
obtained by tuning the relative phase, $\phi$ between the wires.
To further illustrate this control, we performed Rabi oscillations
for various relative phases (Fig. \ref{fig:RabiCircular}). We started
with a parameric scan in the \textbf{weak} driving regime. The NV
spin was driven with both wires, each with amplitude $\Omega=0.15\omega_{L}^{-}$
($\left(2\pi\right)$5.9MHz), and the relative phase between the sources
was scanned (Fig. \ref{fig:RabiCircular}a). At the optimal phase
relation , $\phi=\pi/2$, the NV spin is driven most efficiently,
resulting in Rabi oscillations at double the frequency $0.3\omega_{L}^{-}$
(Fig. \ref{fig:RabiCircular}a, red dashed line), corresponding to
driving with the co-rotating field, and the counter-rotating term
is suppressed completely ($\varepsilon_{-}=1,\varepsilon_{+}=0$).
In contrast, at $\phi=3\pi/2$ the spin remained untouched (Fig. \ref{fig:RabiCircular}a,
green dotted line), as the MW has the opposite polarization to drive
the spin transition. Here, the co-rotating field does not exist ($\varepsilon_{-}=0,\varepsilon_{+}=1$),
and the counter-rotating field can be neglected via the RWA $\left(\Omega\ll\omega_{L}^{+}\right)$.
In contrast, in the \textbf{strong} driving regime the spin is also
driven by the counter-rotating field (Fig. \ref{fig:RabiCircular}b).
Here, we set $\Omega=0.7\omega_{L}^{-}$ for each wire, and only at
$\phi=\pi/2$ (representing left-handed circular polarization, $\sigma^{-}$)
pure harmonic oscillations were observed, demonstrating Rabi flops
with $\Omega=1.4\omega_{L}^{-}$. Hereafter, we denote the ratio of
the Rabi frequency to the Larmor frequency as $\lambda=\Omega/\omega_{L}^{-}$.
For other phases, more complex dynamics were observed, accompanied
with high frequency components and lower contrast, specifically, at
$\phi>\pi$, the dynamics are governed by the counter-propagating
field and one notices an increase in the oscillation frequency with
low contrast.

For both cases, a numerical model based on Eq.(\ref{eq:StrongDrivingHamiltonian})
reproduces the results very well. For all non-circular polarizations,
the phase of the MW with respect to the pulse rising edge has an important
role in the dynamics. For example, with linearly polarized MW radiation,
the effective driving field is $\tilde{\Omega}=\sqrt{2}\Omega e^{-i\left(\omega_{L}^{-}t-\pi/4\right)}\cos\left(\omega_{L}^{-}t-\phi_{g}\right)$,
representing a field with time-dependent magnitude and orientation.
Assuming a square pulse shape (the rise and fall times of the experimental
pulses were $\sim$1ns), one finds that for $\phi_{g}=0$, the field
is $\tilde{\Omega}=\sqrt{2}\Omega\cos$$\left(\omega_{L}^{-}t\right)$,
and both fields (co- and counter-) start with maximal amplitude in
the same direction (in the rotating frame), effectively rotating the
spin instantaneously. In contrast, for $\phi_{g}=\pi/2$, the effective
field is $\tilde{\Omega}=\sqrt{2}\Omega\sin$$\left(\omega_{L}^{-}t\right)$.
Here, the field has zero amplitude at $t=0,$ the spin starts to rotate
much slower, drawing a different trajectory on the Bloch sphere. Conventionally,
and in our experiments too, the trigger of the MW switch is not synchronized
with the MW source phase, leading to a randomized initial phase $\phi_{g}$
over all acquisitions (each sequence was repeated $\sim$10$^{5}$
times for sufficient photon statistics). Therefore, the simulated
signal plotted in Fig. \ref{fig:RabiCircular} is the averaged signal
of 300 repetitions of the dynamics under Eq.(\ref{eq:StrongDrivingHamiltonian})
with uniformly distributed, global phases. In the weak driving regime
the initial microwave phase is unimportant and the repeated acqusitions
are essentially identical. For more details on the global phase dependence,
see Appendix B.

\subsection{Strong driving with linear and circular fields}

After optimizing the relative phase for circular polarization ($\phi=\pi/2$)
and for linear polarization ($\phi=0$), we compare the performance
of the two polarizations for manipulating the spin in the \textbf{strong}
driving regime. Specifically we compare the ability to steer the spin
from the north pole of the Bloch sphere,$\left|0\right\rangle $ to
the south pole, $\left|-1\right\rangle $, i.e. to perform a $\pi$
pulse.

\begin{figure}
\includegraphics[width=1.02\columnwidth]{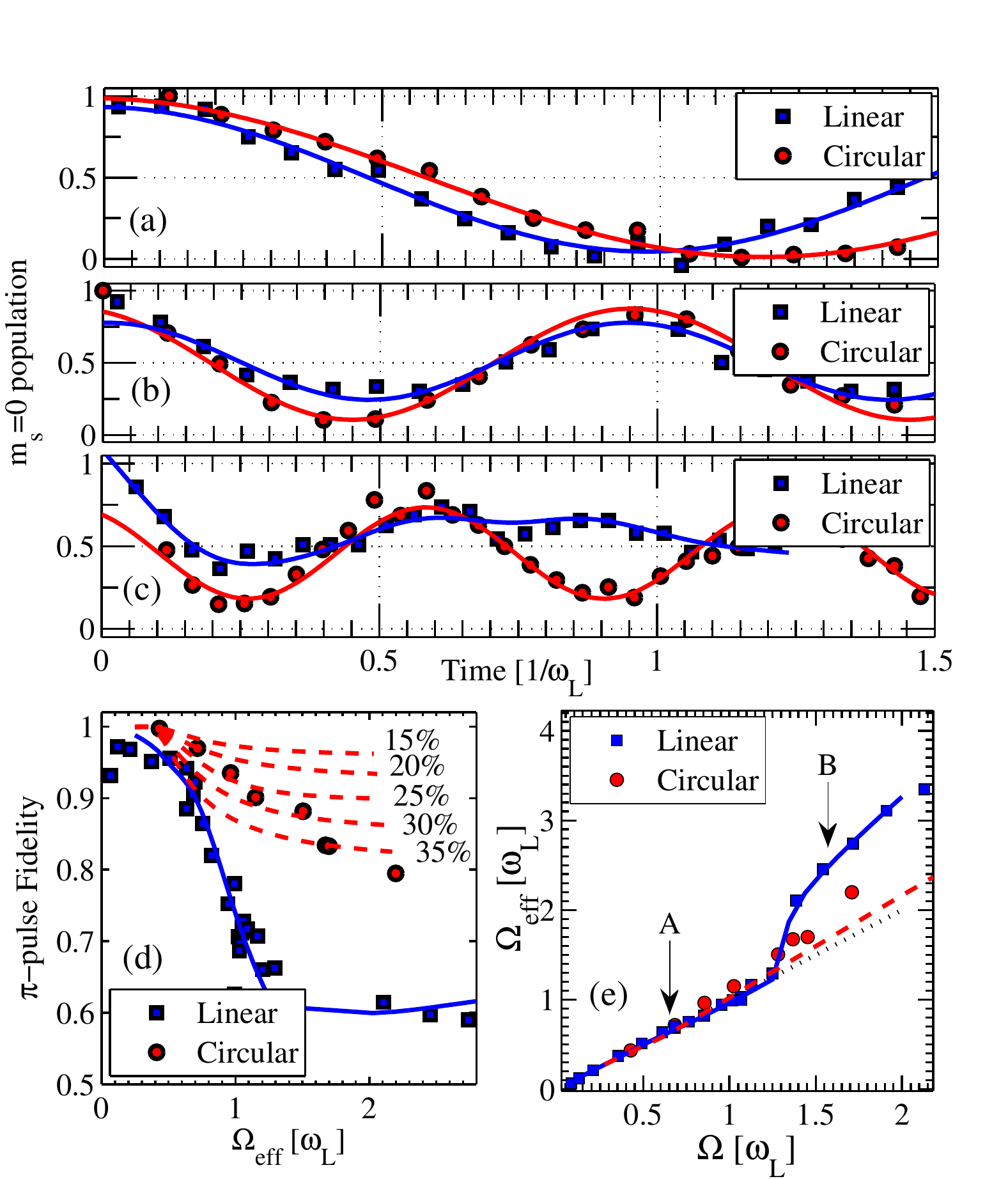}

\caption{Rabi oscillations with linearly and circularly polarized driving fields
(color online). (a -- c) Rabi oscillations for various amplitudes
of MW field: experimental data in rectangles for linear field, and
in circles for circular fields. Data is presented for (a) $\lambda\simeq$0.5,
(b)$\lambda\simeq$1, and (c) $\lambda\simeq$1.5. The solid lines
are cosine fits to the circular polarization data, and simulated signals
from Eq. (\ref{eq:StrongDrivingHamiltonian}) for the linear polarization
data. (d) Experimental comparison of $\pi$ pulse fidelity for linearly
(squares) and circularly (circles) polarized fields. The linear polarization
data is modeled with a curve calculated according to Eq.(\ref{eq:StrongDrivingHamiltonian})
with no fit parameters (solid blue curve). The circular polarzation
data is modeled with an additional axial MW field of amplitude 15\%-35\%$\Omega$
(dashed red curves). (e) The effective Rabi frequency vs. the field
amplitude for linearly (squares) and circularly (circles) polarized
fields. The ideal dependence for harmonic driving is given by the
dotted line. The ``A'' and ``B'' labels correspond to Fig. \ref{fig:Global-phase-dependance.}d
in the Appendix.\label{fig:CircularVsLinear}}
\end{figure}

Fig. \ref{fig:CircularVsLinear}a, b, and c, show the spin dynamics
for $\lambda\sim$0.5, 1.0 and 1.5, respectively. A qualitative difference
is observed in the spin dynamics as the driving field exceeds the
Larmor frequency; the oscillations become anharmonic for linear fields
whilst remaining harmonic for circular fields. We extract two quantities
from the measured signals: the time of the first minimum of the signal,
$t_{m}$, and the $\left|0\right\rangle $-state population at this
time. The former corresponds to a $\pi$-pulse duration for ideal
harmonic driving, and the latter corresponds to the $\pi$-pulse fidelity,
i.e. how well the spin is transferred from the $\left|0\right\rangle $
state to the $\left|-1\right\rangle $ state. 

Fig. \ref{fig:CircularVsLinear}d shows the fidelity of $\pi$-pulse
as a function of the effective Rabi strength (defined as half of the
inverse of the $\pi$-pulse duration, i.e. $\Omega_{eff}=1/2t_{m}$).
For linear polarization the $\pi$-pulse fidelity decreases substantially
when $\lambda\geq1$ (Fig. \ref{fig:CircularVsLinear}d, rectangles),
as predicted by a model based on Schr�dinger equation with Eq. (\ref{eq:StrongDrivingHamiltonian})
(Fig. \ref{fig:CircularVsLinear}d, solid line). In contrast, for
circular polarization the fidelity is 93\% at $\lambda=1$. Importantly,
harmonic behavior of the driven spin is still observed for a Rabi
frequency of twice the Larmor frequency. In principle the fidelity
shouldn't decrease even in the strong driving regime, however, for
high $\lambda$-values the experimental values show monotonic reduction
in the $\pi$-pulse fidelity. This behavior can be partly reproduced
by simulations, if an additional field which is applied parallel to
the NV-axis is included. This is illustrated by the five dashed lines
in Fig. \ref{fig:CircularVsLinear}d, for which we added to Eq. (\ref{eq:StrongDrivingHamiltonian})
an additional term $H_{\parallel}=S_{z}\cdot$$\Omega_{z}\cos$$\left(\omega_{L}^{-}t\right)$,
where $S_{z}$ is the spin operator in the z-axis, and $\Omega_{z}$
is the MW projection on the z-axis. The improved agreement between
experiment and simulation for values of $\Omega_{z}/\Omega$= 20\%-30\%,
implies that this could be a dominant mechanism for the degraded performance
of circularly polarized radiation as the field amplitude is increased.

An additional figure of merit for the manipulation performance is
the how the driving speed changes with the applied microwave amplitude
(Fig. \ref{fig:CircularVsLinear}e). Here, for a linearly polarized
MW field at $\lambda>1.2$ multiple minima appear in the flourescence
signal (see Appendix B). As a consequence, the time of the first minimum
changes abruptly at these values, shifting from the predicted behavior
of $\Omega=\Omega_{eff}=1/2t_{m}$ to higher values (Fig. \ref{fig:CircularVsLinear}e,
rectangles), in an agreement with our numerical model (Fig. \ref{fig:CircularVsLinear}e,
solid line) {[}the ideal behavior is depicted as a dotted black line{]}.
In contrast, for circular polarization the $\pi$-pulse duration follows
the ideal behaior (Fig. \ref{fig:CircularVsLinear}e, circles), with
a small deviation towards higher values. Again, this is partly explained
by including an axial field (red dashed line, calculated with $\Omega_{z}/\Omega$=
20\%).

\section{Discussion}

\begin{figure*}
\includegraphics[width=1\textwidth]{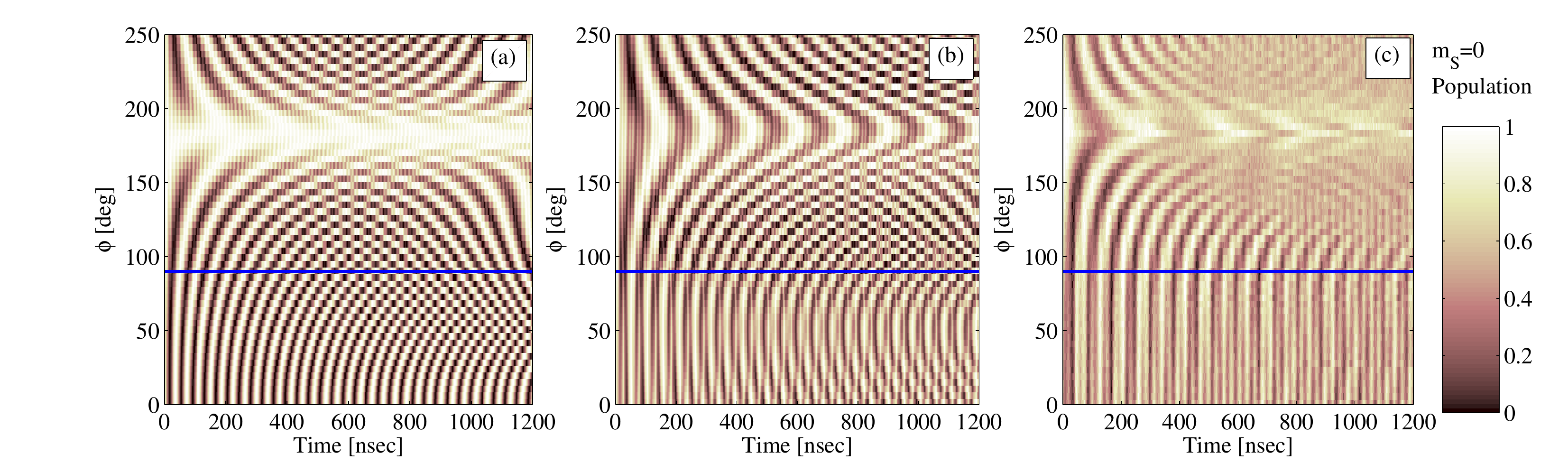}

\caption{Strong driving of an NV center with large axial MW field (color online).
Simulation of Rabi oscillations ($\omega_{L}^{-}=\left(2\pi\right)37.8$MHz,
$\Omega{}_{x}=\Omega_{y}=1.408$$\omega_{L}^{-}$) with (a) $\Omega_{z}=0$
and (b) $\Omega_{z}=1.67\Omega_{x}$. (c) Experimental Rabi oscillations
for a single NV center with $\omega_{L}^{-}$ and $\Omega_{x,y}$
equal to the values in the simulations, and with strong axial component.\label{fig:Z-component}}
\end{figure*}

The comparison of the experiments and simulations in Fig. \ref{fig:CircularVsLinear}
indicates that an axial MW field could have an important role in our
driving scheme. To verify this effect, we measured NV centers at various
positions relative to the cross-wires, and selected an NV center with
high axial component of the MW field. A scan of Rabi flops as a function
of the relative phase between the wires is shown in Fig. \ref{fig:Z-component}.
Here, without an axial component of the MW field, one would expect
to obtain the results in Fig. \ref{fig:Z-component}a, where at $0^{\circ}$
(representing $\sigma^{-}$ polarization) the spin is driven with
harmonic oscillations by the co-rotating field. In contrast, the experiment
shows a qualitatively different behavior, where for phases in the
range $0-90^{\circ}$, the oscillation frequency remains relatively
constant, and the shape is clearly anharmonic (the blue solid line
in Fig. \ref{fig:Z-component} is a guide for the eye). Moreover,
at the cancellation point ($180^{\circ}$), the spin is still rotated
with Rabi frequency about fifth of the applied $\Omega$ (Fig. \ref{fig:Z-component}c).
Remarkably, a numerical simulation based on Eq.(\ref{eq:StrongDrivingHamiltonian})
with axial component $S_{z}\Omega_{z}$, reproduces these features,
elucidating the importance of axial driving for this NV center (Fig.
\ref{fig:Z-component}b). At the cancellation point, for instance,
the spin is likely to be driven via multiple Landau-Zener transitions
\cite{ZhouPRL2014}, rather than with conventional Rabi flops. 

Thus, the performance of our current design of a polarized MW antenna
(cross-wires configuration) in the strong driving regime, is sensitive
to the projection of the MW field on the NV center axis. Although
axial driving is important for realizing Landau-Zener like transitions
\cite{FuchsNaturePhys2011,ChildressPRA2010,ZhouPRL2014}, it is also
accompanied with a reduction of the oscillation contrast.

\section{Conclusion and outlook}

In conclusion, we studied the dynamics of a single spin under resonant,
polarized MW radiation. The relative phase of two MW sources was utilized
as a knob to adjust the MW polarization. We demonstrated high fidelity
selective excitation within a dense spectrum of resonances, allowing
individual excitation of adjacent transitions ($\Delta\sim26$MHz)
with fast pulses of $\Omega\simeq$114MHz. This is of importance near
level crossings, where conventionally one would have to decrease the
driving power to avoid leakage of population to neighboring states,
or use optimal control solutions \cite{DoldeNcomm2014,WaldherrNature2014}.
Here, the selection rules obtained with circularly polarized light
allow selective excitation of degenerate transitions and can be used
to determine both the sign and magnitude of the external magnetic
field \cite{MichlAPL2014}. We showed that under circular MW fields,
the spin experiences pure harmonic oscillations regardless of the
applied field strength, and specifically even above the RWA limit
(in our case more than twice the Larmor frequency). Importantly, although
being in the strong driving regime, the spin is still rotated in its
rotating frame, allowing for universal control around the Bloch sphere
by controlling the global phase of the MW fields $\phi_{g}$. This
enables the use of complex dynamical decoupling schemes\cite{GullionJMR1969},
with sub-Larmor period intrapulse duration. Moreover, in continuous
wave sensing schemes such as dressed-state magnetometry \cite{DegenPRL2013}
and Hartmann-Hahn double resonance \cite{LondonPRL2013}, the spin
must be maintained in its dressed state. Here, our scheme overcomes
the upper limit to detection frequencies set by the Larmor frequency.

Our current design suffers from the influence of an axial component
of the MW field. More versatile stuctures, for example using two wires
for generating each magnetic field $B_{x,y}$ component, could mitigate
this problem by allowing cancellation of the axial field without supressing
the transverse component. Spin manipulation with an axial field is
strongly connected to Landau-Zener transitions \cite{ZhouPRL2014},
and coherent destruction of tunneling \cite{ChildressPRA2010}, and
is therefore interesting in and of itself. Moreover, our ability to
control the magnetic field in all three direction can assist in constructing
the Berry Hamiltonian \cite{Berry1984}, for acquiring controlled
geometric phases with a single spin in diamond without rotating the
sample \cite{MaclaurinPRL2012}.

\bigskip

\bigskip
\begin{acknowledgments}
The authors thank Jochen Scheuer, Xi Kong, and Christoph M�ller for
assistance with experiments. The authors are grateful to Philip Hemmer,
David Gershoni, Ran Fischer and Chen Avinadav for fruitful discussions
and suggestions. The research was supported by DARPA, EU (ERC Synergy
grant BioQ, DIAMANT), DFG (SFB TR 21, FOR 1493, FOR 1482), RSF, the
Alexander von Humboldt and Volkswagen foundations.
\end{acknowledgments}
\bibliographystyle{pnas}
\bibliography{Bibliography,NVreferences}

\section*{Appendix A: Derivation of the Eq. (\ref{eq:FullHamiltonian}) }

We we derive the MW terms in the Hamiltonian of Eq.(\ref{eq:FullHamiltonian}).
The interaction term of the NV spin $\mathbf{S}$, with MW field of
frequency $\omega$, Rabi frequency $\Omega$, and relative phase
$\phi$, is 
\begin{eqnarray*}
H_{MW} & = & \Omega(\cos\left(\omega t\right)S_{x}+\cos\left(\omega t+\phi\right)S_{y})
\end{eqnarray*}
where the transverse spin operators are 
\[
S_{x}=\frac{1}{\sqrt{2}}\left[\begin{array}{ccc}
0 & 1 & 0\\
1 & 0 & 1\\
0 & 1 & 0
\end{array}\right],S_{y}=\frac{1}{\sqrt{2}}\left[\begin{array}{ccc}
0 & -i & 0\\
i & 0 & -i\\
0 & i & 0
\end{array}\right].
\]
We move into the MW rotating frame using the transformation $H^{\prime}=i\frac{dU}{dt}U^{\dagger}-U^{\dagger}HU$
with $U=\exp$$\left(iAt\right)$ and $A$ is given by
\[
A=\left[\begin{array}{ccc}
\omega & 0 & 0\\
0 & 0 & 0\\
0 & 0 & \omega
\end{array}\right].
\]
Then, the transformation operators are
\[
U=\left[\begin{array}{ccc}
e^{i\omega t} & 0 & 0\\
0 & 1 & 0\\
0 & 0 & e^{i\omega t}
\end{array}\right],U^{\dagger}=\left[\begin{array}{ccc}
e^{-i\omega t} & 0 & 0\\
0 & 1 & 0\\
0 & 0 & e^{-i\omega t}
\end{array}\right],
\]
and the spin operators transform according to\begin {widetext}
\begin{eqnarray*}
U^{\dagger}S_{x}U & = & \frac{1}{\sqrt{2}}\left[\begin{array}{ccc}
e^{-i\omega t} & 0 & 0\\
0 & 1 & 0\\
0 & 0 & e^{-i\omega t}
\end{array}\right]\left[\begin{array}{ccc}
0 & 1 & 0\\
1 & 0 & 1\\
0 & 1 & 0
\end{array}\right]\left[\begin{array}{ccc}
e^{i\omega t} & 0 & 0\\
0 & 1 & 0\\
0 & 0 & e^{i\omega t}
\end{array}\right]=\frac{1}{\sqrt{2}}\left[\begin{array}{ccc}
0 & e^{-i\omega t} & 0\\
e^{i\omega t} & 0 & e^{i\omega t}\\
0 & e^{-i\omega t} & 0
\end{array}\right]\\
U^{\dagger}S_{y}U & = & \frac{1}{\sqrt{2}}\left[\begin{array}{ccc}
e^{-i\omega t} & 0 & 0\\
0 & 1 & 0\\
0 & 0 & e^{-i\omega t}
\end{array}\right]\left[\begin{array}{ccc}
0 & -i & 0\\
i & 0 & -i\\
0 & i & 0
\end{array}\right]\left[\begin{array}{ccc}
e^{i\omega t} & 0 & 0\\
0 & 1 & 0\\
0 & 0 & e^{i\omega t}
\end{array}\right]=\frac{1}{\sqrt{2}}\left[\begin{array}{ccc}
0 & -ie^{-i\omega t} & 0\\
ie^{i\omega t} & 0 & -ie^{i\omega t}\\
0 & ie^{-i\omega t} & 0
\end{array}\right].
\end{eqnarray*}
The terms in the Hamiltonian are calculated as follows
\begin{eqnarray*}
\Omega\cos\left(\omega t\right)U^{\dagger}S_{x}U & = & \Omega\frac{1}{2}\left(e^{i\omega t}+e^{-i\omega t}\right)\frac{1}{\sqrt{2}}\left[\begin{array}{ccc}
0 & e^{-i\omega t} & 0\\
e^{i\omega t} & 0 & e^{i\omega t}\\
0 & e^{-i\omega t} & 0
\end{array}\right]=\frac{\Omega}{2\sqrt{2}}\left[\begin{array}{ccc}
0 & 1+e^{-2i\omega t} & 0\\
1+e^{2i\omega t} & 0 & 1+e^{2i\omega t}\\
0 & 1+e^{-2i\omega t} & 0
\end{array}\right]\\
\cos\left(\omega t+\phi\right)U^{\dagger}S_{y}U & = & \Omega\frac{1}{2}\left(e^{i\left(\omega t+\phi\right)}+e^{-i\left(\omega t+\phi\right)}\right)\frac{1}{\sqrt{2}}\left[\begin{array}{ccc}
0 & -ie^{-i\omega t} & 0\\
ie^{i\omega t} & 0 & -ie^{i\omega t}\\
0 & ie^{-i\omega t} & 0
\end{array}\right]\\
 & = & \frac{\Omega}{2\sqrt{2}}\left[\begin{array}{ccc}
0 & -i\left(e^{i\phi}+e^{-i\left(2\omega t+\phi\right)}\right) & 0\\
i\left(e^{i\left(2\omega t+\phi\right)}+e^{-i\phi}\right) & 0 & -i\left(e^{i\left(2\omega t+\phi\right)}+e^{-i\phi}\right)\\
0 & i\left(e^{i\phi}+e^{-i\left(2\omega t+\phi\right)}\right) & 0
\end{array}\right],
\end{eqnarray*}
and combine the two terms, we write the Hamiltonian in the MW rotating
frame $H'_{MW}=U^{\dagger}\left[\Omega\cos\left(\omega t\right)S_{x}+\cos\left(\omega t+\phi\right)S_{y}\right]U$
: 
\begin{eqnarray*}
H'_{MW} & = & \frac{\Omega}{2\sqrt{2}}\left[\begin{array}{ccc}
0 & 1+e^{-2i\omega t} & 0\\
1+e^{2i\omega t} & 0 & 1+e^{2i\omega t}\\
0 & 1+e^{-2i\omega t} & 0
\end{array}\right]\\
 & + & \frac{\Omega}{2\sqrt{2}}\left[\begin{array}{ccc}
0 & -i\left(e^{i\phi}+e^{-i\left(2\omega t+\phi\right)}\right) & 0\\
i\left(e^{i\left(2\omega t+\phi\right)}+e^{-i\phi}\right) & 0 & -i\left(e^{i\left(2\omega t+\phi\right)}+e^{-i\phi}\right)\\
0 & i\left(e^{i\phi}+e^{-i\left(2\omega t+\phi\right)}\right) & 0
\end{array}\right]\\
 & = & \frac{\Omega}{\sqrt{2}}\left[\begin{array}{ccc}
0 & \varepsilon_{-} & 0\\
\varepsilon_{-}^{*} & 0 & \varepsilon_{+}\\
0 & \varepsilon_{+}^{*} & 0
\end{array}\right]+\frac{\Omega}{\sqrt{2}}\left[\begin{array}{ccc}
0 & e^{-2i\omega t}\varepsilon_{+} & 0\\
e^{2i\omega t}\varepsilon_{+}^{*} & 0 & e^{2i\omega t}\varepsilon_{-}\\
0 & e^{-2i\omega t}\varepsilon_{-}^{*} & 0
\end{array}\right]
\end{eqnarray*}
where in the last row we assigned $\varepsilon_{\pm}=e^{\mp i\phi_{g}}\left(1-ie^{\mp i\phi}\right)/2$.
\end {widetext}.

\section*{Appendix B: Global phase dependence in the linear polarization data}

\begin{figure}
\includegraphics[width=1\columnwidth]{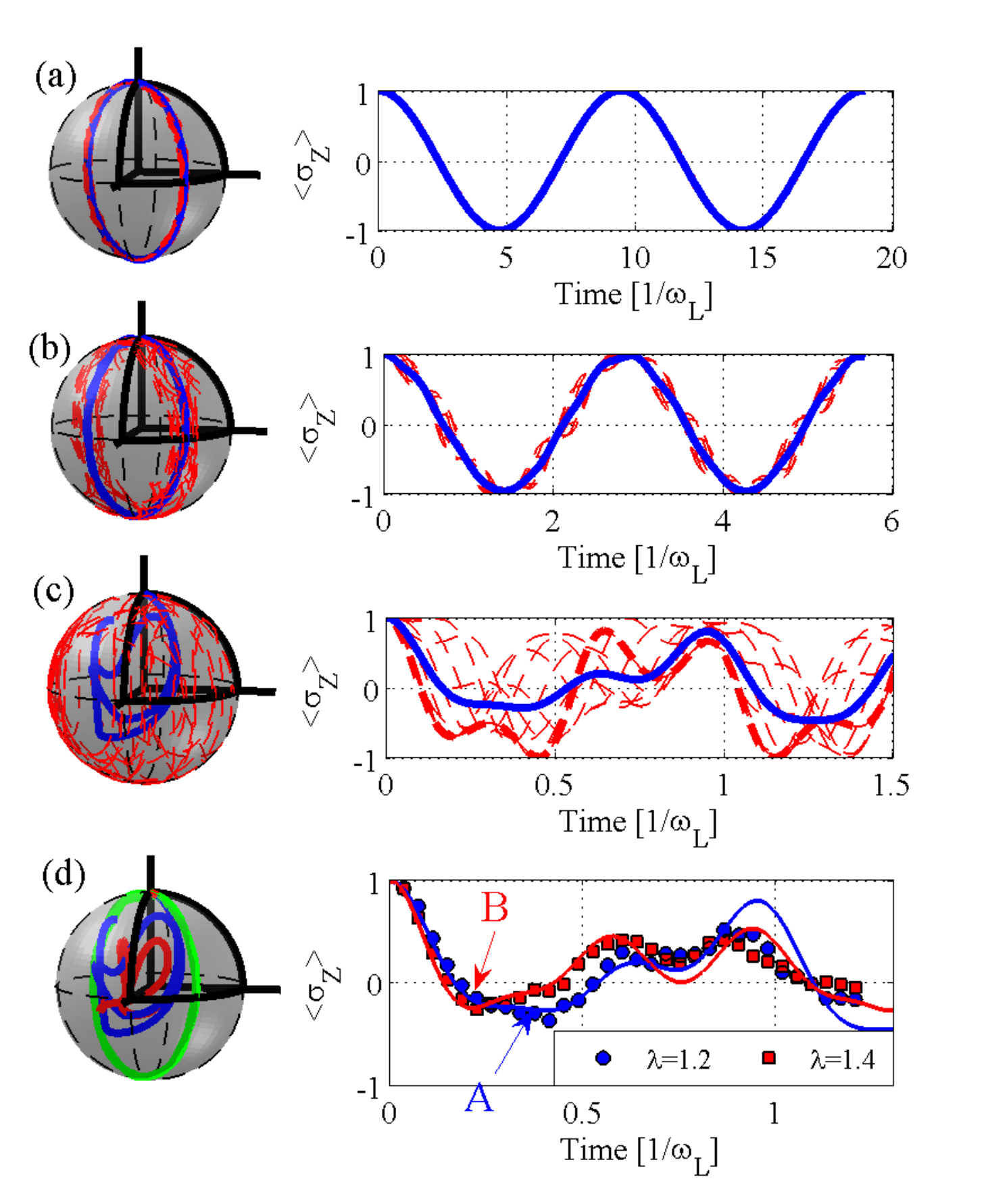}

\caption{Global phase dependance in the strong driving regime. (a-c) Rabi oscillations
for increasing Rabi-frequency values, $\lambda=0.1$(a), $\lambda=0.33$(b),
$\lambda=1.2$(c). On the left - Bloch sphere representation of the
spin state. On the right - the spin's z-component as a function of
the driving duration. Red dashed curves are various realziations of
$\phi_{g}$, and the solid blue curve is an averaged trajectory over
300 realizations. (d) Comparison between experimental values (circles,
rectangels), and numerical simulations for high $\lambda$ values
(1.2, and 1.4 resepectively), the green curve is of Rabi oscillations
at the weak driving regime and serves as guide for the eyes. \label{fig:Global-phase-dependance.}}
\end{figure}

In the main text, Fig. \ref{fig:CircularVsLinear}d,e present the
analysis of Rabi oscillations with linear polarized MW field: the
population transfer at the first minimum point in the signal decreases
substantially as the Rabi frequency increases, and the $\pi$-pulse
duration (the evolution time until the first minimum) changes when
the RWA is exceeded. In contrast, in ref. \cite{FuchsScience2009}
it is found that close-to-unity population transfer can occur also
in the strong driving regime, and that the $\pi$-pulse duration is
very hard to predict. The contridiction arises from the experimental
technique which was used in our work namely to average many realizations
of the MW phase at the pulse rising edge, $\phi_{g}$. In \cite{FuchsScience2009,AvinadavArXiv2014,ScheuerArXiv2014},
the MW phase was syncronized to the pulse edge. While our technique
supresses the sensitivity to imperfections in the driving system,
it occumpanies a systematic reduction in the driving perofmances.
It worth mentioning that there is no systematic deterioration when
applying strong \textbf{circular} MW fields. Here we describe the
global-phase dependence using numerical simulations, compared with
the measured Rabi oscillations signals. Fig. \ref{fig:Global-phase-dependance.}a-c
show the associated dynamics for various field strengths, and for
various $\phi_{g}$-values. At $\lambda=0.1$ (Fig. \ref{fig:Global-phase-dependance.}a),
the influence of the counter rotating term is neglibgle and the dynamics
is identical for any $\phi_{g}$ at the appropriated rotating frame
(red-dashed curves). Then, the spin rotates around a big circle on
the Bloch sphere. At $\lambda=0.33$ (Fig. \ref{fig:Global-phase-dependance.}b),
the dynamics changes for each $\phi_{g}$, but the averaged time trace
still resembles harmonic oscillations (blue solid curve). This demonstrates
the robustness of the averaging technique. At higher Rabi frequency
$\lambda=1.2$ (Fig. \ref{fig:Global-phase-dependance.}c), the counter
rotating term influences the spin rotations markedly and for each
$\phi_{g}$ the dynamics is complete different; while for a given
$\phi_{g}$ complete population transfer from $\left\langle \sigma_{Z}\right\rangle =+1$
to $\left\langle \sigma_{Z}\right\rangle =-1$ can occur (Fig. \ref{fig:Global-phase-dependance.}c,
red thick dashed curve), for the ensemble-average of many $\phi_{g}$,
the population transfer doesn't exceed 60\%, demonstrating the downside
of this technique. When averaging many $\phi_{g}$ realizations, the
principle minimum point of the signal, which marks the $\pi$-pulse
operation, changes into multiple minima structure, as shown in Fig.
\ref{fig:Global-phase-dependance.}d. The circles and rectangles are
experimental values measured at $\lambda=1.2,1.4$, respectively.
The solid lines are the results of a numerical simulation averaging
300 different $\phi_{g}$-values. From the Bloch sphere representation
it is clear that at these values the averaged spin-state becomes mixed
(red, blue curves) compared to the rotation at the weak driving regime
(green curve). In Fig. \ref{fig:Global-phase-dependance.}d, we have
marked the points ``A'' and ``B'' which are adressed in the main
text and in Fig. \ref{fig:CircularVsLinear}.

\end{document}